\documentclass[amsmath,amssymb,aps,prl,reprint,showpacs,floats,superscriptaddress,longbibliography]{revtex4-2}
    \usepackage[utf8]{inputenc}
    \usepackage[table,xcdraw]{xcolor}
    \usepackage{graphicx}
    \usepackage{amsmath, amsthm, amssymb, amsfonts}
    \usepackage{mathtools}
    \usepackage[colorlinks=true,citecolor=blue,linkcolor=blue,urlcolor=blue]{hyperref}
    \usepackage{footnote}
    \usepackage{xifthen}    
    \usepackage{capt-of}
    \usepackage{subfigure}
    \usepackage{physics}
    \usepackage{adjustbox}
    \pdfoutput=1 
    \newcommand{\KDQ}{KDQ }

    \newcommand{\cbqty}[1]{\left\{#1\right\}}

    \newcommand{\sn}[1]{\mathfrak{#1}}
    \newcommand{\opr}[1]{#1}

    \newcommand{\HH}{\opr{H}}
    \newcommand{\VV}{\opr{V}}
    \newcommand{\UU}{\opr{U}}
    \newcommand{\SSop}{\opr{S}}
    \newcommand{\bbi}{\opr{\mathbb{I}}}

    \newcommand{\oA}{\mathcal{O}^A}
    \newcommand{\oB}{\mathcal{O}^B}

    \newcommand{\pA}[1][]{\ifthenelse{\equal{#1}{}}{\mathcal{A}}{\mathcal{A}^{(#1)}}}
    \newcommand{\pB}[1][]{\ifthenelse{\equal{#1}{}}{\mathcal{B}}{\mathcal{B}^{(#1)}}}
    \newcommand{\pC}[1][]{\ifthenelse{\equal{#1}{}}{\mathcal{C}}{\mathcal{C}^{(#1)}}}

    \newcommand{\ssalabel}{\sn{A}}
    \newcommand{\ssblabel}{\sn{B}}
    \newcommand{\ssclabel}{\sn{C}}

    \newcommand{\piA}{\pi^{\ssalabel}}
    \newcommand{\piB}{\pi^{\ssblabel}}

    \newcommand{\bbiA}{\bbi^\ssalabel}
    \newcommand{\bbiB}{\bbi^\ssblabel}
    \newcommand{\bbiC}{\bbi^\ssclabel}

    \newcommand{\HHA}{\HH^\ssalabel}
    \newcommand{\HHB}{\HH^\ssblabel}
    \newcommand{\HHC}{\HH^\ssclabel}
    \newcommand{\HHAB}{\HH^{\ssalabel\ssblabel}}
    \newcommand{\HHAC}{\HH^{\ssalabel\ssclabel}}
    \newcommand{\HHBC}{\HH^{\ssblabel\ssclabel}}
    \newcommand{\VVA}{\VV^\ssalabel}
    \newcommand{\VVB}{\VV^\ssblabel}
    \newcommand{\SSA}{\SSop^\ssalabel}
    \newcommand{\SSB}{\SSop^\ssblabel}

    \newcommand{\csop}[1]{\mathbb{C}_{#1}}

\begin{document}
\title{Compatibility of quantum measurements and the emergence of classical objectivity}
\author{Emery~Doucet}
\email{emery.doucet@umbc.edu}
\affiliation{Department of Physics, University of Maryland, Baltimore County, Baltimore, MD 21250, USA}
\author{Sebastian~Deffner}
\email{deffner@umbc.edu}
\affiliation{Department of Physics, University of Maryland, Baltimore County, Baltimore, MD 21250, USA}
\affiliation{National Quantum Laboratory, College Park, MD 20740, USA}
\date{\today}
\begin{abstract}
The study of measurements in quantum mechanics exposes many of the ways in which the quantum world is different.
For example, one of the hallmarks of quantum mechanics is that observables may be incompatible, implying among other things that it is not always possible to find joint probability distributions which fully capture the joint statistics of multiple measurements. 
Instead, one must employ more general tools such as the Kirkwood-Dirac quasiprobability (KDQ) distribution, which may exhibit negative or non-real values heralding non-classicality. 
In this Letter, we consider the \KDQ distributions describing arbitrary collections of measurements on disjoint components of some generic multipartite system. 
We show that the system dynamics ensures that these distributions are classical if and only if the Hamiltonian supports Quantum Darwinism. 
Thus, we demonstrate a fundamental relationship between these two notions of classicality and their emergence in the quantum world.
\end{abstract}
\maketitle
%

%
%%%%%%%%%%%%%%%%%%%%%%%%%%%%%%%%%%%%%%%%%%%%%%%%%%%%%%%%%%%%%%%%%%%%%%%
%%%%%% INTRODUCTION
%%%%%%%%%%%%%%%%%%%%%%%%%%%%%%%%%%%%%%%%%%%%%%%%%%%%%%%%%%%%%%%%%%%%%%%
%

The rules of quantum mechanics which describe the universe at a fundamental level are quite different than the classical physics that governs the behavior of our everyday experiences. 
In particular, understanding how an observer performing measurements on a quantum system learns information about that system has a number of complexities and subtleties which are absent for a classical observer. 
This raises an important question: if the universe is fundamentally quantum, why does it usually appear classical? 
That is, how exactly does classicality emerge in quantum systems?

One of the most powerful approaches to the study of the emergence of classicality is to take an intuitive property of classical observations, turn that property into a precise statement, and finally determine how and when that property emerges in quantum systems.
Most famously, the study of classical objectivity -- the fact that many observers can agree on features they obtain from observations of a system -- and its emergence in quantum systems is at the core of Quantum Darwinism \cite{Zurek03, Zurek09, BlumeKohout05, BlumeKohout06, Ollivier04, Ollivier05, Riedel2010PRL, Riedel2011NJP, Brandao2015NC, Knott2018PRL, Milazzo2019PRA, Colafranceschi2020JPA, Touil22b, Girolami2022PRL, Zurek22, QDAndFriends, Zurek2000, Zurek03}.
The key insight is that in practice any observer measuring a quantum system does so indirectly, by inferring information about the system from some fragment of the environment with which the system interacts. 
Due to the system's interaction with the environment, it generally decoheres \cite{BreuerPetruccione}.
In models exhibiting Quantum Darwinism, this decoherence is associated with the redundant encoding of information about the system in the so-called pointer basis into the environment \cite{Zurek81, Zurek82, Zurek03, Zurek09, Zurek22, Ollivier04, Ollivier05, Riedel2011NJP, Riedel2010PRL}, which allows many observers to learn the same information about the system. 

It has recently been shown \cite{Touil24} that there is a unique structure of system-environment states which support Quantum Darwinism: the state must be of ``singly-branching form'' \cite{BlumeKohout06, Zurek22, Korbicz21} (equivalently, the system and environment must support a ``spectrum broadcast structure'' \cite{Korbicz21, Le19, Horodecki15, Korbicz17}).
Given this fact, it is possible to provide explicit classifications of system-environment Hamiltonians that support Quantum Darwinism \cite{Duruisseau23, Doucet24}.
In other words, it is possible to start with a statement of a behavioral difference between quantum and classical measurements and develop a classification of exactly which quantum models behave classically.

\begin{figure}
    \centering
    \includegraphics[width=0.73\columnwidth]{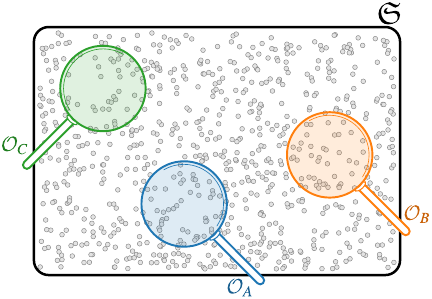}
    \caption{Distinct observers $\mathcal{O}_{A, B, C}$ measuring some large composite quantum system $\sn{S}$ often measure disjoint collections of degrees of freedom (gray dots), with much of the system remaining inaccessible. The notion of classical compatibility we study in this work is related to if and when the joint statistics of these distinct measurements can be captured with a classical probability distribution, independent of the precise details of the measurements themselves.}
    \label{fig:sketch}
\end{figure}

In this Letter, we replicate this success starting from another difference between quantum and classical measurements: \emph{all classical measurements are compatible, whereas not all quantum measurements are.}
If multiple observers perform independent measurements of some classical system, it is always possible to construct a joint probability distribution which faithfully captures all the features of the statistics of the observers' measurements. 
This is true regardless of the precise details of the measurements or of the system state.
With a quantum system this is generally not true, even if the system is a large composite system and the observers measure independent fragments, as illustrated in Fig.~\ref{fig:sketch}.
Non-commutativity in quantum mechanics means that measurements may be incompatible, and hence it is generally not possible to construct joint probability distributions for quantum measurements \cite{BreuerPetruccione, Lostaglio23}.
Instead, the measurement statistics are described by joint quasiprobability distributions which may take negative or complex values.

Perhaps the most famous such quasiprobability distribution is the Wigner distribution \cite{Wigner1932, Tatarski83}, a joint distribution for position and momentum important to phase space formulations of quantum mechanics.
For more general observables, e.g.,
\begin{align}
    \oA = \sum_{i} a_i \pA_i
    \quad\textrm{and}\quad
    \oB = \sum_{j} b_j \pB_j
    ,
    \label{eqn:ProjDecompAB}
\end{align}
where $\pA_i, \pB_j$ are projectors corresponding to the different measurement outcomes, one can use the Kirkwood-Dirac quasiprobability (KDQ) distribution \cite{Kirkwood1933, Dirac1945} to describe the joint measurement statistics.
The quasiprobability of the outcome $\oA_i$ and $\oB_j$ is simply,
\begin{align}
    q_{ij} = \tr\bqty{\pB_j\pA_i\rho}
    .
\end{align}
Note that if the projectors commute, this quasiprobability is in fact just the probability of measuring outcome $\pA_i$ followed by $\pB_j$ in a two-point measurement (TPM) protocol,
\begin{align}
    p_{ij}^{\rm TPM} = \tr\bqty{\pB_j\pA_i\rho\pA_i}
    .
\end{align}
In general, each quasiprobability may be written as a combination of the TPM probability and two quantum modification terms \cite{He24, Johansen07}.

The KDQ description of joint measurements has seen a tremendous amount of interest in recent years \cite{ArvidssonShakur24, Lostaglio23} with a broad range of applications, e.g., in tomography and understanding weak values \cite{Lostaglio23, ArvidssonShakur24}, in characterizing quantum work distributions and other facets of quantum thermodynamics \cite{HernandezGomez24, Mazzola13, Lostaglio23, ArvidssonShakur24, Halpern17, Gherardini24}, and in understanding many-body systems and out-of-time-ordered correlators \cite{ArvidssonShakur24, Lostaglio23, Gherardini24}.
Most relevant to this work, it also has applications in quantum foundations and in characterizing non-classicality in quantum measurements \cite{Schmid24, He24, Johansen07, Gherardini24, ArvidssonShakur24}. 
Negative or non-real values of the KDQ distribution are useful as a measure of non-classicality \cite{He24, Lostaglio23, Alonso19, Budiyono24}, and can help witness quantum contextuality \cite{ArvidssonShakur24, Schmid24}.

\par
\textit{Classical compatibility}.--%
Inspired by the operational definition of classical objectivity in Quantum Darwinism, we define ``classical compatibility'' of measurements: in a classical system, any set of measurements performed by a collection of observers can be captured with a joint probability distribution, independent of the system state.
In a quantum system, classical compatibility emerges if the same is true of any collection of quantum measurements performed on it.

When no restrictions are placed on the allowed sets of measurements, classical compatibility can never emerge in a quantum system as it is always possible to find incompatible measurements \footnote{For example, select any two non-degenerate eigenstates. Take the system state to be one of them and consider two measurements to be performed immediately after one another, one diagonal and one not. These measurements are incompatible and so the KDQ distribution for this scenario does not coincide with the TPM distribution.}.
If as in Fig.~\ref{fig:sketch} we instead suppose that each measurement acts on a subsystem and that these subsystems are disjoint (i.e., we assume each observer to have exclusive access to whatever subsystem they measure), then as we will show there do exist quantum systems which exhibit emergent classical compatibility. 
This is a property of the model and Hamiltonian describing its time-evolution; clearly simultaneous measurements of disjoint subsystems are always compatible, but if there is a delay between the measurements then the dynamics determines whether or not the measurements remain compatible. 

In the following, we identify which models support this emergent notion of classicality, with exactly the same generality as true classical systems: even when each observers' choice of measurement and measurement time is completely free, the joint statistics should be describable with a classical joint probability distribution.
In the remainder of this Letter we will give a full characterization of exactly which quantum models have this property, and will show that they are in fact exactly the same models which exhibit Quantum Darwinism \cite{Doucet24}.
This is despite the fact that we have made no reference to any of the key information-theoretic concepts central to Quantum Darwinism, only to a simple property of the joint statistics of classical measurements. 

%
%%%%%%%%%%%%%%%%%%%%%%%%%%%%%%%%%%%%%%%%%%%%%%%%%%%%%%%%%%%%%%%%%%%%%%%
%%%%%% PROOF
%%%%%%%%%%%%%%%%%%%%%%%%%%%%%%%%%%%%%%%%%%%%%%%%%%%%%%%%%%%%%%%%%%%%%%%
%
\par
\textit{Characterization}.--%
Consider the scenario of Fig.~\ref{fig:sketch}, where some composite system $\sn{S}$ is made available to a set of observers $\{\oA, \oB, \dots\}$, where each observer has exclusive access to some collection of subsystems.
Each observer specifies a measurement they wish to perform, e.g., specified by sets of projectors $\{\pA_i\}$, $\{\pB_j\}$, and a measurement time.
The system is subject to some time-independent Hamiltonian $\HH$, which describes how the system evolves between measurements. 

If that particular collection of measurements exhibits no KDQ non-classicality \cite{Lostaglio23, He24, Alonso19}, then they are jointly measurable and can be captured with a classical joint probability distribution.
If \emph{all} sets of disjoint measurements given \emph{all} initial system states $\rho_0$ can be described by classical joint probability distributions, then the observers cannot possibly witness non-classical correlations or effects between their respective degrees of freedom -- the system appears classical to them from the point of view of joint measurement statistics.

We begin with the simplest case of two observers who perform projective measurements at times $t_A < t_B$ of the observables defined in Eq.~\eqref{eqn:ProjDecompAB}, with the individual projectors evolving in the Heisenberg picture as $\pA_i(t) = \UU^\dagger(t) \pA_i \UU(t)$ where $\UU(t) = \exp(-it\HH)$ is the propagator.
Note that we are not limited by considering projective measurements, as each observer may perform POVMs by expanding the system to include additional degrees of freedom accessible only to them on which their generalized measurement is projective \cite{NielsenChuang}.

Denoting the inter-measurement time $\tau \equiv t_B - t_A$, the associated \KDQ distribution reduces to the TPM distribution and so the measurement statistics can be described with a joint probability distribution regardless of the initial system state $\rho_0$ if and only if \cite{Lostaglio23},
\begin{align}
    \comm{\pA_i}{\pB_j(\tau)} = \comm{\pA_i}{\UU^\dagger(\tau) \pB_j \UU(\tau)} = 0 \qquad \forall i,j
    .
    \label{eqn:ABComm}
\end{align}
Further, since classical compatibility requires that this is true for any choice of observables $\oA,\oB$ we must ultimately require that this commutator vanishes for any pair of projectors $\pA_i, \pB_j$.
We therefore drop the subscripts moving forward.

Clearly, since the measurements are on different subsystems, if the measurements are simultaneous ($\tau = 0$) this commutator is always zero.
When $\tau \ne 0$, this commutator is guaranteed to be zero only if the dynamics supports no mechanism for back-action between the two measurements.
This is only possible if the Hamiltonian $\HH$ has a specific form.

Using the Baker-Campbell-Hausdorff formula we write the commutator of Eq.~\eqref{eqn:ABComm} as an infinite series of nested commutators. 
Defining $\pB[n]$ as
\begin{align}
    \pB[n] = \comm{\HH}{\pB[n-1]}, \quad\textrm{with}\quad \pB[0] = \pB
    ,
\end{align}
then
\begin{align}
    \comm{\pA}{\UU^\dagger(\tau)\pB\UU(\tau)}
        &= \sum_{n=0}^{\infty} \frac{(i\tau)^n}{n!} \comm{\pA}{\pB[n]}
    .
    \label{eqn:BCH}
\end{align}
This sum must vanish termwise, so we will proceed term-by-term to determine what Hamiltonians force each nested commutator to be zero. 
The $n=0$ term is always zero since the projectors act on different subsystems, so $n=1$ gives the first non-trivial term.

With two observers, we split the system $\sn{S}$ into three components: the subsystems accessible to the first observer $\sn{A}$, those accessible to the second $\sn{B}$, and any inaccessible to either observer $\sn{C}$.
Using this decomposition, the two observers' projectors in the Schr\"odinger picture have the form,
\begin{align}
    \pA = \piA \otimes \bbiB \otimes \bbiC
    \quad\textrm{and}\quad
    \pB = \bbiA \otimes \piB \otimes \bbiC
    .
\end{align}
and the general form of the full tripartite Hamiltonian is,
\begin{align}
    \HH = \sum_{ijk} h_{ijk} \pqty{P_i \otimes Q_j \otimes S_k}
\end{align}
where $P_i, Q_j, S_k$ are operators drawn from some orthogonal basis of operators on $\sn{A}, \sn{B}$, and $\sn{C}$, respectively.

If for each $k$ we let $\HHAB_k = \sum_{ij}h_{ijk}\pqty{P_i\otimes Q_j}$, then the commutator in the $n=1$ term of Eq.~\eqref{eqn:BCH} is
\begin{align}
    \comm{\pA}{\comm{\HH}{\pB}}
    &= \sum_{k} \comm{\piA \otimes \bbiB}{\comm{\HHAB_k}{\bbiA \otimes \piB}} \otimes S_k
    .
\end{align}
Since the $S_k$ are orthogonal this sum must vanish termwise, which is only possible to guarantee for all $\piA,\piB$ if each $\HHAB_{k}$ is a sum of local terms \footnote{Consider the operator Schmidt decomposition of the Hamiltonian. If there are components of the form $P \otimes Q$ representing interactions, then we may choose projectors $\pA$ and $\pB$ which fail to commute with those interactions. Intuitively, the issue is that interacting Hamiltonians lead to entanglement, which can affect certain measurements leading to differences  between the KDQ and TPM distributions.}
That is, the $n=1$ term of Eq.~\eqref{eqn:BCH} is guaranteed to be zero if and only if the Hamiltonian is of the form,
\begin{align}
    \HH = \HHA + \HHB + \HHC + \HHAC + \HHBC
    .
    \label{eqn:HForm}
\end{align}
Note that this implies that the bipartite case where there is no inaccessible subsystem $\sn{C}$ is highly restricted.
In that case, classical compatibility requires the Hamiltonian to be of the form $\HHA + \HHB$, corresponding to two uncoupled systems. 

Now, we turn to the $n=2$ term of Eq.~\eqref{eqn:BCH}.
Since $\pB[1] = \comm*{\HHB + \HHBC}{\pB}$ only has support on subsystems B and C, we have
\begin{align}
    \pB[2] = \comm{\HHB + \HHC + \HHAC + \HHBC}{\pB[1]}
    .
    \label{eqn:B2Rec}
\end{align}
Only the term with $\HHAC$ can have support on subsystem $\sn{A}$, so the $n=2$ term becomes
\begin{align}
    \comm{\pA}{\pB[2]} &= \comm{\pA}{\comm{\HHAC}{\pB[1]}}
    .
\end{align}
For this to vanish for all $\pA$, it must be ensured that $\comm*{\HHAC}{\pB[1]} = 0$.
From Eq.~\eqref{eqn:B2Rec}, we see that this would also ensure that $\pB[2]$ only has support on subsystems $\sn{B}$ and $\sn{C}$.
We may repeat the same analysis for the $n=3$ term, where we find that we must require that $\comm*{\HHAC}{\pB[2]} = 0$, which forces $\pB[3]$ to have support only on subsystems $\sn{B}$ and $\sn{C}$, and so on.
Starting from Eq.~\eqref{eqn:HForm}, each term ($n\ge2$) of the expansion in Eq.~\eqref{eqn:BCH} is guaranteed to be zero for all $\pA$ if and only if $\comm*{\HHAC}{\pB[n-1]} = 0$.

Note that $\HHAC$ and $\pB[n-1]$ can only fail to commute due to their actions on subsystem $\sn{C}$.
Suppose we decompose the interaction Hamiltonians as,
\begin{align}
    \HHAC &= \sum_{a} \VVA_a \otimes \bbiB \otimes \SSA_a
    ,
    \\
    \HHBC &= \sum_{b} \bbiA \otimes \VVB_b \otimes \SSB_b
    ,
\end{align}
such that the operators $\VVA_a$ are orthogonal when $a\ne a'$, and similarly for $\VVB_b$.
Then, for any $n$, substituting these decompositions into $\comm*{\HHAC}{\pB[n-1]}$ yields expressions of the form
\begin{align}
    \sum_a \VVA_a \otimes f(\pB,\HHB,\{\VVB\}) \otimes \comm{\SSA_a}{g(\HHC, \{\SSB\})}
    ,
\end{align}
where $f$ and $g$ represent complicated nested commutator expressions of their arguments. 
Since the $V^A_a$ operators are orthogonal, for the sum to vanish each term must be zero.  
This can be guaranteed for all $\pB$ only if $\comm{\SSA_a}{g(\HHC, \cbqty{\SSB})}$ is zero. 
Note that this is only a constraint on the operators which act on the inaccessible subsystem $\sn{C}$, and that neither the local Hamiltonians $\HHA, \HHB$ nor the action of the interaction Hamiltonians $\HHAC, \HHBC$ on the subsystems $\sn{A}$ and $\sn{B}$ matter. 
Different operators acting on those accessible subsystems change which measurements may witness non-classicality, but not it can be witnessed at all.

Through this procedure, the first few terms in the expansion of Eq.~\eqref{eqn:BCH} translate to the commutator constraints,
\begin{subequations}
\begin{align}
    \comm{\SSA_a}{\SSB_i} &= 0 \qquad \forall a,i
    ,
    \\
    \comm{\SSA_a}{\comm{\HHC}{\SSB_i}} &= 0  \qquad \forall a,i
    ,
    \\
    \comm{\SSA_a}{\comm{\HHC}{\comm{\HHC}{\SSB_i}}} &= 0  \qquad \forall a,i
    ,
    \\
    \comm{\SSA_a}{\comm{\HHC}{\comm{\SSB_j}{\SSB_i}}} &= 0  \qquad \forall a,i,j
    ,
\end{align}
\end{subequations}
with the $n=2$ term giving the single commutator, the $n=3$ term the double commutator, and the $n=4$ term the two triple commutators.
Note that we have assumed the constraints from the $n=2$ term are satisfied when computing the $n=3$ constraints, and so on. 
Otherwise there would be additional redundant constraints for higher terms, e.g., fully expanding the $n=4$ term produces the two triple commutators from above as well as contributions involving single and double commutators, all of which vanish if the lower order constraints are satisfied.

To compactly represent these constraints, we introduce notation for the commutator superoperators,
\begin{align}
    \csop{0}O = \comm{H_C}{O}
    ,
    \qquad
    \csop{i}O = \comm{\SSB_i}{O}
    .
\end{align}
Then, all the commutator constraints we have found can be expressed as,
\begin{align}
    \comm{\SSA_a}{\pqty{\prod_{n=1}^{|\mu|} \csop{\mu(n)}}\SSB_b} = 0
    ,
    \label{eqn:CommConstraint}
\end{align}
where we arbitrarily define the product as growing to the left $(\dots\csop{\mu(2)}\csop{\mu(1)})$ and where $\mu$ represents a sequence with elements in $\cbqty{0,\dots,|\mathcal{L}(C)|}$. 

Every finite sequence $\mu$ corresponds to a nested commutator, and every such commutator must be zero for all $a$ and $b$ to ensure there are no effective interactions between subsystems A and B.
Some sequences generate commutators which are either trivially zero (e.g., a sequence of all zeros) or commutators which are guaranteed to be zero by a constraint from a shorter sequence. For $n=2$, sequences with one positive element give the non-trivial constraints. For $n>2$, new non-trivial constraints are generated by sequences with last element 0 and at least one positive element. 

These commutator constraints are our primary result, as they completely characterize the space of quantum Hamiltonians which support classical compatibility of disjoint measurements. 
Intuitively, they have a simple interpretation: interactions between subsystems $\sn{A}$ and $\sn{B}$ provide a channel for measurement back-action, and it is always possible to choose a pair of measurements with projectors which are sensitive to this and are hence rendered incompatible.
Crucially, this remains true for effective high-order interactions -- the measurements may only be slightly incompatible and hence the timescales at which non-classicality manifests may be long, but the KDQ distributions will eventually become non-classical. 
This can only be avoided if no interactions are present, even effectively, which requires that all the nested commutators of Eq.~\eqref{eqn:CommConstraint} vanish.

%
%%%%%%%%%%%%%%%%%%%%%%%%%%%%%%%%%%%%%%%%%%%%%%%%%%%%%%%%%%%%%%%%%%%%%%%
%%%%%% EXTENSIONS
%%%%%%%%%%%%%%%%%%%%%%%%%%%%%%%%%%%%%%%%%%%%%%%%%%%%%%%%%%%%%%%%%%%%%%%
%
\par
\textit{Many-observer case}.--%
These results may be extended to the case of three or more observers by considering the extension of the \KDQ distribution to the joint description of many measurements.
The only modification to Eq.~\eqref{eqn:CommConstraint} is that $\SSA_a, \SSB_b$ are replaced with operators corresponding to any two observers' subsystems, and that the nested commutators in the product may contain operators from any subsystem (i.e., the elements of the sequence $\mu$ are now pairs $(\sn{B}, i)$ including the subsystem and index).
The intuition for this case is a simple generalization of the two-observer case: the measurement statistics of any collection of measurements on disjoint subsystems can always be described with a joint probability distribution (i.e., the system exhibits classical compatibility) if and only if the Hamiltonian neither contains nor effectively generates interactions between any subsystems.

For example, consider again the scenario of Fig.~\ref{fig:sketch} with three observers performing measurements at different times $t_A = 0$, $t_B = \tau$, and $t_C = \tau'$ on subsystems $\sn{A}$, $\sn{B}$, and $\sn{C}$, and with the remaining subsystems being inaccessible to any observer.
The quasiprobability for the outcome corresponding to the projectors $\pA_i, \pB_j, \pC_k$ is $q_{ijk} = \tr\bqty{\pC_k(\tau')\pB_j(\tau)\pA_i(0)\rho_0}$.
If the Hamiltonian satisfies the commutator constraints and so generates no effective interactions between the subsystems the observers have access to, then all pairs of measurements are compatible and so we may commute $\pA_i(0)$ through the other time-evolved projectors and incorporate it into a modification of the initial state, $\rho_0 \to \tilde{\rho}_i = \pA_i \rho_0\pA_i$. 
We are left with the two-measurement \KDQ joint quasiprobability of the B and C measurements with an initial state which depends on the $i$ index, $q_{ijk} = (\tilde{q}_i)_{jk} \equiv \tr\bqty{\pC_k(\tau')\pB_j(\tau)\tilde{\rho}_i(0)}$, which due to the absence of effective interactions is guaranteed to be an ordinary probability.

%
%%%%%%%%%%%%%%%%%%%%%%%%%%%%%%%%%%%%%%%%%%%%%%%%%%%%%%%%%%%%%%%%%%%%%%%
%%%%%% DARWINISM
%%%%%%%%%%%%%%%%%%%%%%%%%%%%%%%%%%%%%%%%%%%%%%%%%%%%%%%%%%%%%%%%%%%%%%%
%
\par
\textit{Connection to Quantum Darwinism}.--%
For models which exhibit classical compatibility with an arbitrary number of observers who access arbitrary (but still disjoint) collections of subsystems, the criteria we have derived in Eq.~\eqref{eqn:CommConstraint} coincide exactly with the criteria that were found to be necessary for a Hamiltonian to support Quantum Darwinism \cite{Duruisseau23, Doucet24}.
In the context of Quantum Darwinism, the different observers shown in Fig.~\ref{fig:sketch} are measuring independent fragments of an environment. 
Information about some system of interest (which would live in the component of $\sn{S}$ inaccessible to any observer) is redundantly encoded into the environment such that each observer can reconstruct the projection of the system onto each of the various pointer states, themselves selected by the specific system-environment interaction encoded in the Hamiltonian.
The need for unidirectional information transfer to build redundancy for Quantum Darwinism places mild additional restrictions on the Hamiltonian and initial system-environment state such that the mutual information between the system and environment fragments eventually develops a classical plateau \cite{Zurek22, BlumeKohout06, Riedel12, Touil24}. 
The Hamiltonian structure is identical, however.

While it is not surprising that the Hamiltonians that support Quantum Darwinism and classical compatibility are related -- in both cases we look for the emergence of classicality in a quantum model -- it is surprising that they are the \emph{same}. 
Quantum Darwinism is deeply grounded in information theory, and is concerned with what exactly different observers can learn from their measurements.
This contrasts with the notion of classical compatibility, which makes no reference to what if anything the observers learn from their measurements, only to the notion that they should be able to use classical probabilities to describe the joint statistics of their measurements.
The key connection is that both notions of classicality require that the measured degrees of freedom must be non-interacting, even effectively. 
In Quantum Darwinism, the only system-environment states which support objectivity are states of singly-branching form \cite{Touil24}, and such states are destroyed by intra-environment interactions. 
In the present work, we find that interactions between observed components allows the existence of incompatible measurements.

One interesting consequence of the connection we have found is that our results point to a method of testing whether a particular model can exhibit Quantum Darwinism which avoids complicated optimization procedures necessary to determine quantum discord.
Take any two fragments of the environment and consider any two measurements on those fragments. 
If the \KDQ distribution exhibits any non-classicality, from our results it is certain that the model cannot exhibit Quantum Darwinism. 
Actually measuring the complete \KDQ distribution is non-trivial, as complete knowledge of the \KDQ distribution can be equivalent to reduced state tomography \cite{Johansen07,He24,Schmid24,ArvidssonShakur24}.
Therefore, in practice one would measure only a few quasiprobabilities or one of the quantities which can be derived from the KDQ distribution's myriad applications.
How exactly the tests are performed changes the details of the hypothesis testing procedure and of how confidence scales with the number of tests, but the core idea remains the same.

\acknowledgments{S.D. acknowledges support from the John Templeton Foundation under Grant No. 62422.}

\bibliography{references}
\end{document}